      [1999/12/01 v1.4c Il Nuovo Cimento]
\documentclass[preprint]{cimento}

\usepackage{cite}
\usepackage{cancel}
\usepackage{amsmath}
\usepackage{graphicx}
\newcommand{\eq}[1]{Eq.~(\ref{#1})}
\newcommand{\ps}{p\hspace{-0.44em}/\hspace{0.06em}}
\newcommand{\Msusy}{M_{\textrm{SUSY}}}

\newcommand{\gev}{\, \mathrm{GeV}}
\newcommand{\tev}{\, \mathrm{TeV}}

\title{Higgs mediated Flavour Violation in 2HDMs and the MSSM -- An Overview}
\author{A.~Crivellin\from{ins:x}}
\instlist{\inst{ins:x} CERN Theory Division, CH-1211 Geneva 23, Switzerland}
\PACSes{14.65.Fy,14.80.Da,14.80.Ly}
\begin{document}

\maketitle

\begin{abstract}
In these proceedings we review the flavour phenomenology of two-Higgs-doublet models (2HDMs) and connect the results to the decoupling limit of the MSSM. We first study the impact of FCNC constraints on the allowed parameter space of the 2HDM and examine how recent deviations from the SM expectations in tauonic $B$ decays (observed by BABAR) can be explained in a 2HDM with generic flavour structure (of type III) with sizable flavour violation in the up-sector~\cite{Crivellin:2012ye,Crivellin:2013wna}. Afterwards, we discusses the matching of the MSSM on the 2HDM of type III. Here we focus on the two-loop SQCD corrections to the Higgs-quark-quark couplings~\cite{Crivellin:2012zz}.
\end{abstract}

\section{Introduction}
\begin{figure}[htbp]
\begin{center}
\includegraphics[width=0.49\textwidth]{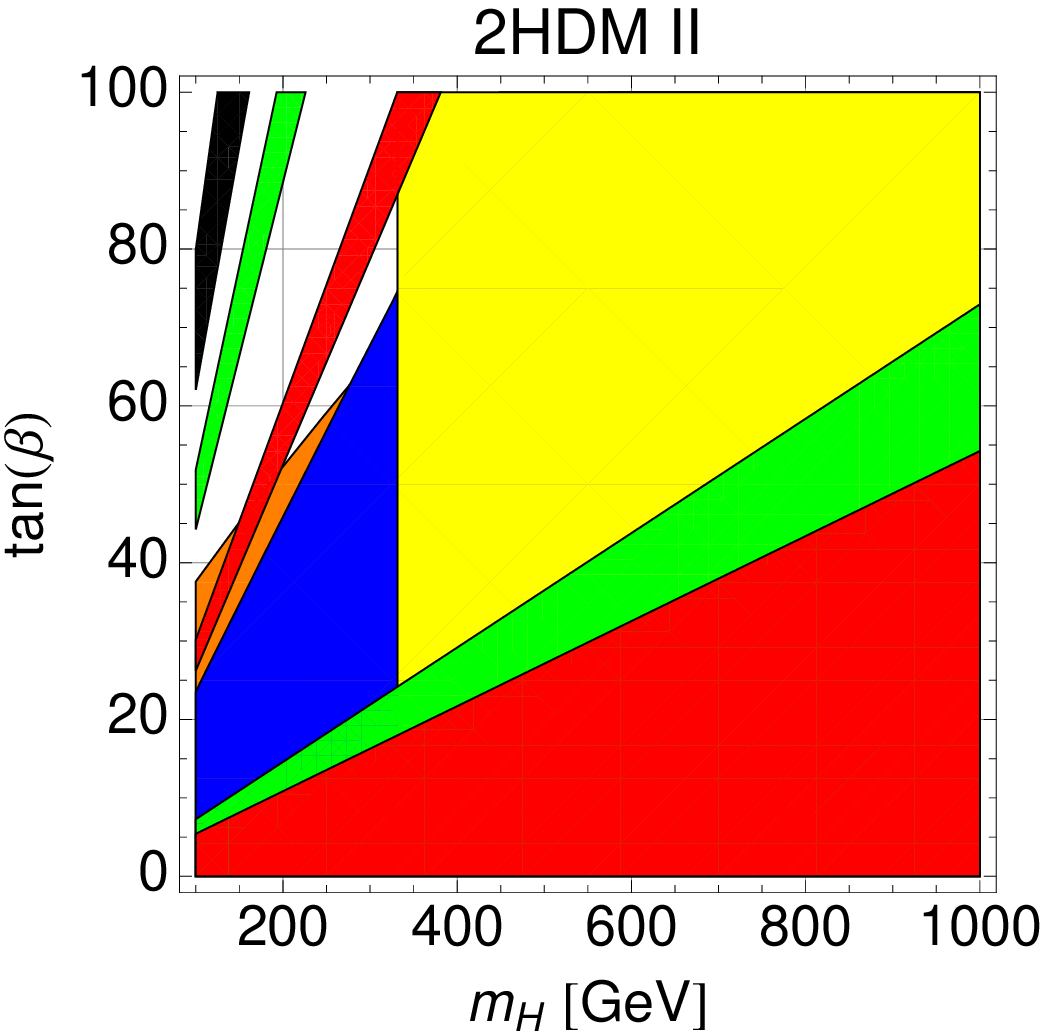}
\includegraphics[width=0.5\textwidth]{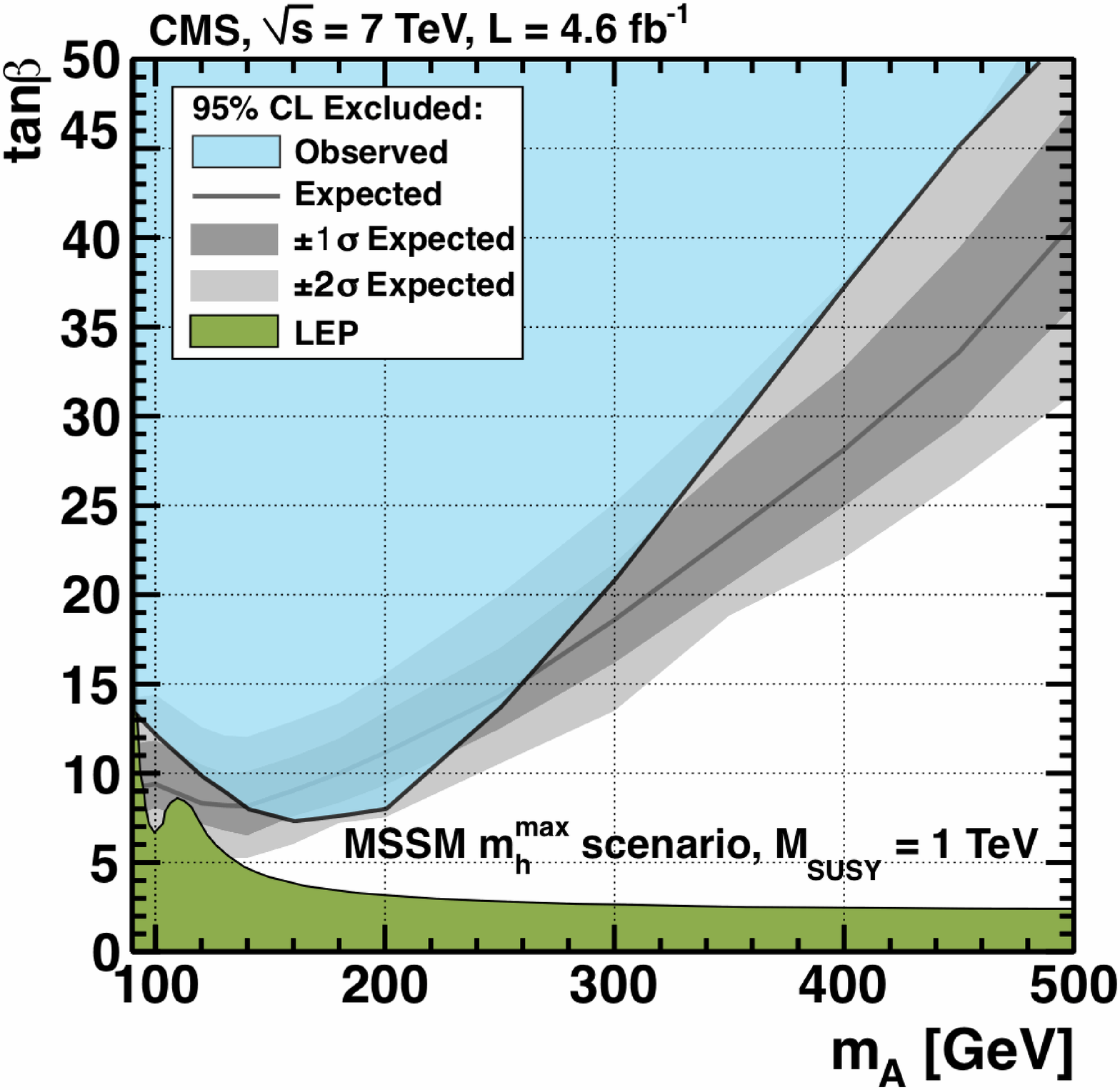}
\end{center}
\caption{Left: Updated constraints on the 2HDM of type II parameter space. The regions compatible with experiment are shown (the regions are superimposed on each other): $b\to s\gamma$ (yellow) \cite{Hermann:2012fc}, $B\to D\tau\nu$ (green), $B\to \tau \nu$ (red), $B_{s}\to \mu^{+} \mu^{-}$ (orange), $K\to \mu \nu/\pi\to \mu \nu$ (blue) and $B\to D^*\tau \nu$ (black). Note that no region in parameter space is compatible with all processes. Explaining $B\to D^*\tau \nu$ would require very small Higgs masses and large values of $\tan\beta$ which is not compatible with the other observables. To obtain this plot, we added the theoretical uncertainty linear on the top of the $2 \, \sigma$ experimental error.\newline
Right: Plot from the CMS collaboration taken from Ref.~\cite{CMS}: Exclusion limits in the $m_{A^0}$--$\tan\beta$ plane from $A^{0}\to \tau^{+}\tau^{-}$. The analysis was done in the MSSM, but since we consider a 2HDM with MSSM-like Higgs potential and the MSSM corrections to the $A^0\tau\tau$ vertex are small, we can apply this bound to our model. However, a large value of $\epsilon^\ell_{33}$ in the 2HDM of type~III could affect the conclusions. Note that in the limit $v\ll m_H$ all heavy Higgs masses ($m_{H^0}$, $m_{A^0}$ and $m_{H^\pm}$) are approximately equal.}
\label{fig:2HDMII}
\end{figure}

The Standard Model (SM) contains only one scalar isospin doublet, the Higgs doublet. After electroweak symmetry breaking, this gives masses to up quarks, down quarks and charged leptons. The charged component of this doublet becomes the longitudinal component of the $W$ boson and the neutral CP-odd component becomes the longitudinal component of the $Z$ boson. Thus we have only one physical neutral Higgs particle. In a 2HDM \cite{Lee:1973iz} we introduce a second Higgs doublet and obtain four additional physical Higgs particles (in the case of a CP conserving Higgs potential): the neutral CP-even Higgs $H^0$, a neutral CP-odd Higgs $A^0$ and the two charged Higgses $H^{\pm}$. The most general Lagrangian for the Yukawa interactions (which corresponds to the 2HDM of type III) in the physical basis with diagonal quark mass matrices is given by
\begin{eqnarray}
\renewcommand{\arraystretch}{2.2}
\begin{array}{l}
\mathcal{L}^{eff}  = \bar u_{f\;L}^{} V_{fj} \left( {\dfrac{{m_{d_i} }}{{v_d }}\delta_{ij}H_d^{2\star}  - \epsilon_{ji}^{d} \left( {H_u^1  + \tan \left( \beta  \right)H_d^{2\star} } \right)} \right) d_{i\;R}  \\ 
\phantom{\mathcal{L}^{eff}  =}  
+ \bar d_{f\;L} V_{j f}^{\star} \left( {\dfrac{{m_{u_j} }}{{v_u }}\delta_{ij}H_u^{1\star}  - \epsilon_{ji}^{u} \left( {H_d^2  + \cot \left( \beta  \right)H_u^{1\star} } \right)} \right) u_{i\;R}  \\ 
\phantom{\mathcal{L}^{eff}  =}  
- \bar d_{f\;L}  \left( {\dfrac{{m_{d_i} }}{{v_d }}\delta_{fi}H_d^{1\star}  + \epsilon_{fi}^{d} \left( {H_u^2  - \tan \left( \beta  \right)H_d^{1\star} } \right)}  \right) d_{i\;R}  \\ 
\phantom{\mathcal{L}^{eff}  =}  - \bar u_{f\;L}^a \left( {\dfrac{{m_{u_i} }}{{v_u }}\delta_{fi}H_u^{2\star}  + \epsilon_{fi}^{u} \left( {H_d^1  - \cot \left( \beta  \right)H_u^{2\star} } \right)} \right)u_{i\;R} \,+\,h.c.  \\ 
 \end{array}
\label{L-Y-FCNC}
\end{eqnarray}
where $\epsilon^q_{ij}$ parametrizes the non-holomorphic corrections which couple up (down) quarks to the down (up) type Higgs doublet\footnote{Here the expression ``non-holomorphic" already implicitly refers to the MSSM where non-holomorphic couplings involving the complex conjugate of a Higgs field are forbidden due to the holomorphicity of the superpotential.}. In the MSSM at tree-level $\epsilon^q_{ij}=0$, which also corresponds to the 2HDM of type II, flavour changing neutral Higgs couplings are absent ($\epsilon^q_{ij}=0$). A combination of flavour constraints on the 2HDM of type II is given in the left plot of Fig.~\ref{fig:2HDMII}.

However, at the loop-level, the non-holomorphic couplings $\epsilon^q_{ij}$ are generated~\cite{Hamzaoui:1998nu}\footnote{See the second article of Ref.~\cite{Crivellin:2010er} for a complete treatment of all chirally enhanced effects.} and in the following we will assume that $\epsilon^q_{ij}$ are free parameters but are small corrections compared to the Yukawa coupling, i.e. $|v_u\epsilon^d_{ij}\leq {\rm{max}[m_{d_i}m_{d_j}]}|$ and $|v_d\epsilon^u_{ij}\leq {\rm{max}[m_{u_i}m_{u_j}]}|$ which is in agreement with 't Hooft's naturalness criterion.

\section{Constraints from FCNC processes}

\subsection{Tree-level constraints}
\begin{figure}[t]
\centering
\includegraphics[width=0.3\textwidth]{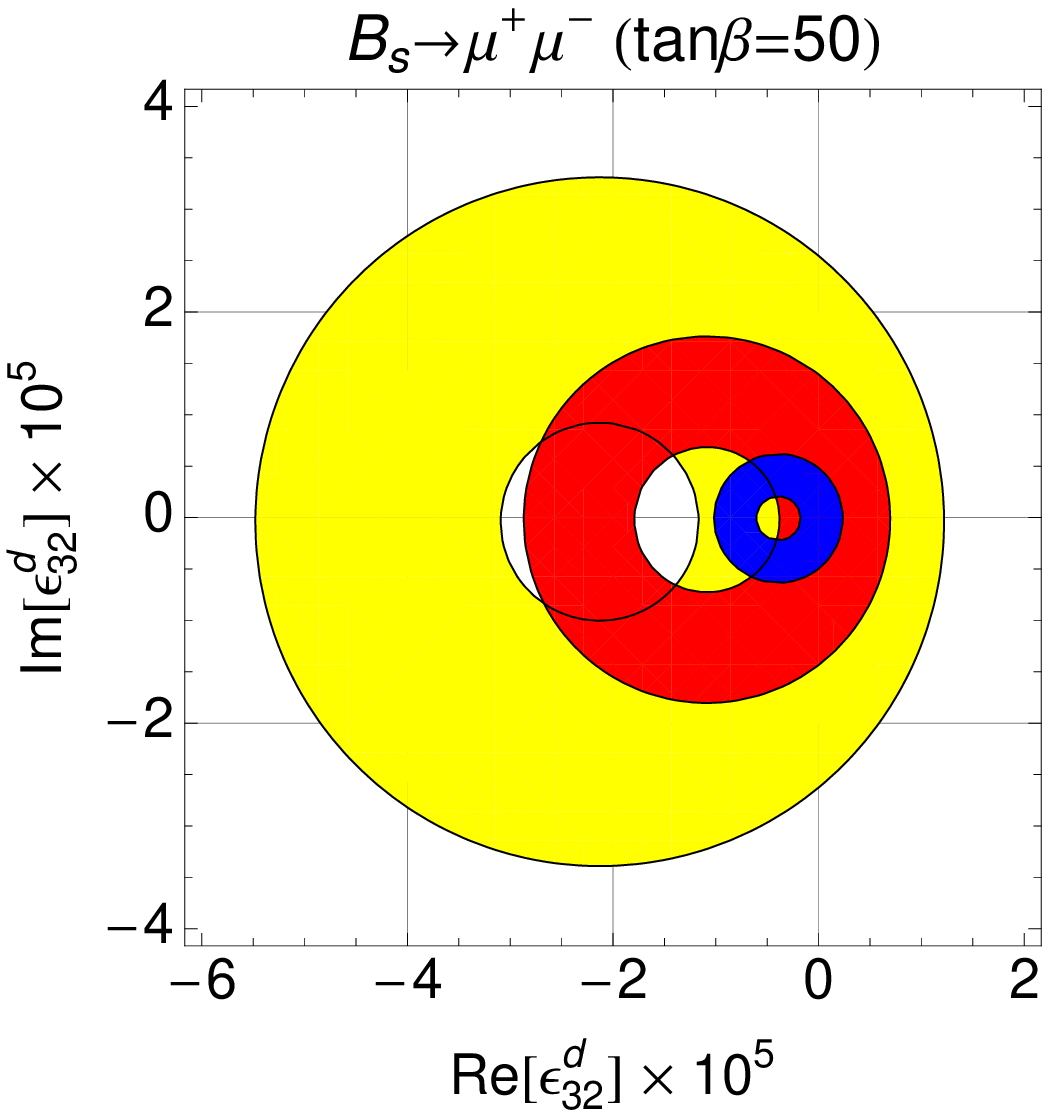}~~~~~
\includegraphics[width=0.3\textwidth]{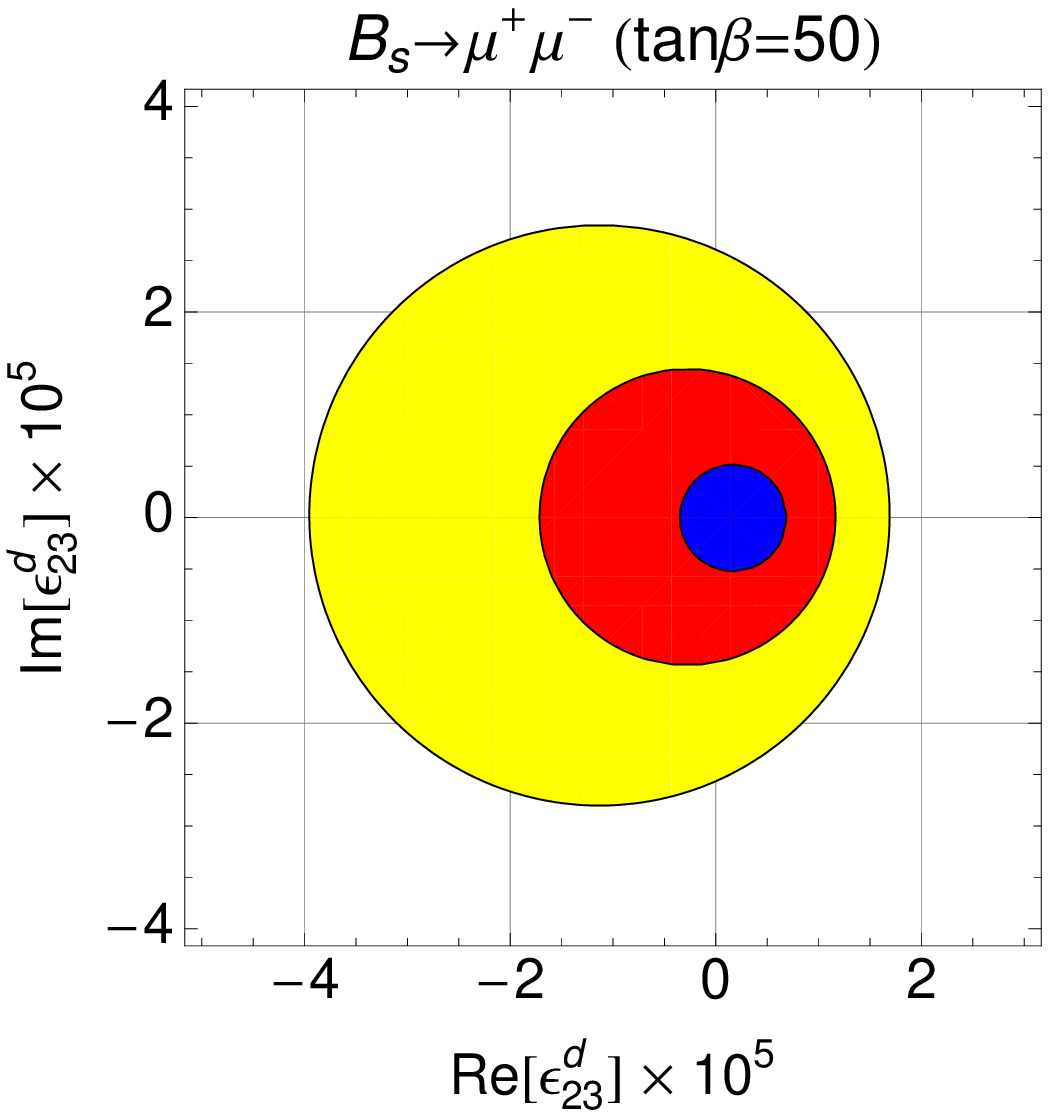}
\includegraphics[width=0.334\textwidth]{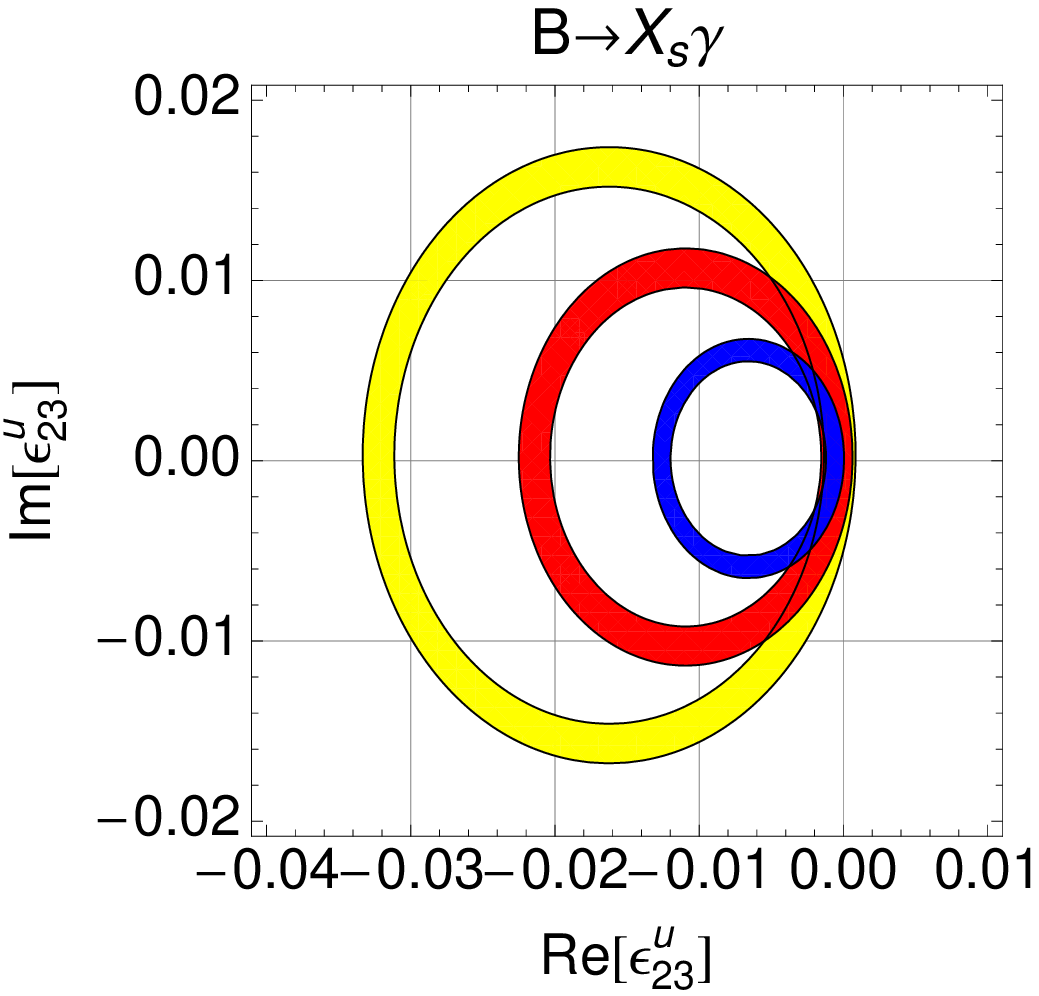}
\caption{Left (middle): Allowed regions in the complex $\epsilon^{d}_{23(32)}$--plane from $B_s\to\mu^+\mu^-$ for $\tan\beta=50$ and $m_{H}=700\mathrm{~GeV}$ (yellow), $m_{H}=500\mathrm{~GeV}$ (red) and $m_{H}=300\mathrm{~GeV}$ (blue). Note that the allowed regions for $\epsilon^{d}_{32}$--plane are not full circles because in this case a suppression of ${\cal B}\left[B_{s}\to\mu^+\mu^-\right]$ below the experimental lower bound is possible.\newline
Right: Allowed regions for $\epsilon^{u}_{23}$ from $ B \to X_{s} \gamma$, obtained by adding the $2\,\sigma$ experimental error and theoretical uncertainty linear for $\tan\beta=50$ and $m_{H}=700 \, \mathrm{ GeV}$ (yellow), $m_{H}=500\, \mathrm{ GeV}$ (red) and  $m_{H}=300 \,\mathrm{ GeV}$ (blue). }
\label{fig:Bstomumu}
\end{figure}

Direct constraints on the off-diagonal elements $\epsilon^q_{fi}$ can be obtained from neutral Higgs contributions to the leptonic neutral meson decays ($B_{s,d}\to\mu^+\mu^-$, $K_L\to\mu^+\mu^-$ and ${\bar D}^0\to\mu^+\mu^-$) which arise already at the tree level \cite{Sher:1991km}\footnote{In principle, the constraints from these processes could be weakened, or even avoided, if $\epsilon^\ell_{22}\approx m_{\ell_{2}}/v_u$. Anyway, in here we will assume that the Peccei-Quinn breaking for the leptons is small and neglect the effect of $\epsilon^\ell_{22}$ in our numerical analysis for setting limits on $\epsilon^q_{ij}$.
}. $K_L\to\mu^+\mu^-$ constrains $\left|\epsilon^d_{12,21}\right|$, $D^0\to\mu^+\mu^-$ imposes bounds on $\left|\epsilon^u_{12,21}\right|$ and $B_s\to\mu^+\mu^-$ ($B_d\to\mu^+\mu^-$) limits the possible size of $\left|\epsilon^d_{23,32}\right|$ $\left(\left|\epsilon^d_{13,31}\right|\right)$. We find the following (approximate) bounds on the absolute value of $\epsilon^q_{ij}$:
\begin{equation}
\renewcommand{\arraystretch}{1.4}
\begin{array}{l}
\left|\epsilon^d_{12,21}\right|\leq 1.6\times 10^{-6}\,,\qquad
\left|\epsilon^u_{12,21}\right|\leq 3 \times 10^{-2}\,,\\
\left|\epsilon^d_{23,32}\right|\leq 3 \times 10^{-5}\,,\qquad
\left|\epsilon^d_{13,31}\right|\leq 1 \times 10^{-5}\,,\\
\end{array}
\end{equation}
for $\tan\beta=50$ and $m_H=500$~GeV. As an example we show the full dependence of the constraints in the complex $\epsilon^{d}_{23,32}$-plane from $B_s\to\mu^+\mu^-$ in left and middle plot of Fig.~\ref{fig:Bstomumu}. Note that both an enhancement or a suppression of ${\cal B}\left[ B_{d,s}\to\mu^+\mu^-\right]$ compared to the SM prediction is possible. If at the same time both elements $\epsilon^{d}_{23}$ and $\epsilon^{d}_{32}$ are non-zero, constraints from $B_s$ mixing arise which are even more stringent.
\medskip

\subsection{Loop constraints}
So far we were able to constrain all flavour off-diagonal elements $\epsilon^d_{ij}$ and $\epsilon^u_{12,21}$ but no relevant tree-level constraints on $\epsilon^u_{13,31}$ and $\epsilon^u_{23,32}$ can be obtained due to insufficient experimental data for top FCNCs. Nonetheless, it turns out that also the elements $\epsilon^u_{13,23}$ can be constrained from charged Higgs contributions to the radiative $B$ decay $b\to d \gamma$ and $ b\to s \gamma$. As an example we show the constraints on $\epsilon^u_{23}$ in the right plot of Fig.~\ref{fig:Bstomumu}. The constraints on $\epsilon^u_{13}$ from $ b\to d \gamma$ are even more stringent \cite{Crivellin:2011ba}.

However, there are no relevant constraints on $\epsilon^u_{32,31}$ from FCNC processes because of the light charm or up quark propagating in the loop (which also requires the contribution to be proportional to this small mass). This has important consequences for charged current processes (to be studied in the next section) where these elements enter.

\section{Tauonic $B$ decays in the 2HDM of type III}

Tauonic $B$-meson decays are an excellent probe of new physics: they test lepton flavor universality satisfied in the SM and are sensitive to new particles which couple proportionally to the mass of the involved particles (e.g. Higgs bosons) due to the heavy $\tau$ lepton involved. Recently, the BABAR collaboration performed an analysis of the semileptonic $B$ decays $B\to D\tau\nu$ and $B\to D^*\tau\nu$ using the full available data set \cite{BaBar:2012xj}. They find for the ratios
\begin{equation}
{\cal R}(D^{(*)})\,=\,{\cal B}(B\to D^{(*)} \tau \nu)/{\cal B}(B\to D^{(*)} \ell \nu)\,,
\end{equation}
the following results:
\begin{eqnarray}
{\cal R}(D)\,=\,0.440\pm0.058\pm0.042  \,,\\
{\cal R}(D^*)\,=\,0.332\pm0.024\pm0.018\,.
\end{eqnarray}
Here the first error is statistical and the second one is systematic. Comparing these measurements to the SM predictions
\begin{eqnarray}
{\cal R}_{\rm SM}(D)\,=\,0.297\pm0.017 \,, \\
{\cal R}_{\rm SM}(D^*) \,=\,0.252\pm0.003 \,,
\end{eqnarray}
we see that there is a discrepancy of 2.2\,$\sigma$ for $\cal{R}(D)$ and 2.7\,$\sigma$ for $\cal{R}(D^*)$ and combining them gives a $3.4\, \sigma$ deviation from the SM~\cite{BaBar:2012xj}. This evidence for new physics in $B$-meson decays to taus is further supported by the measurement of $B\to \tau\nu$ 
\begin{equation}
{\cal B}[B\to \tau\nu]=(1.15\pm0.23)\times 10^{-4}\,.
\end{equation}
which disagrees with by $1.6\, \sigma$ higher than the SM prediction using $V_{ub}$ from a global fit of the CKM matrix \cite{Charles:2004jd}.

A natural possibility to explain these enhancements compared to the SM prediction is a charged scalar particle which couples proportionally to the masses of the fermions involved in the interaction: a charged Higgs boson. A charged Higgs affects $B\to \tau\nu$~\cite{Hou:1992sy}, $B\to D\tau\nu$ and $B\to D^*\tau\nu$~\cite{Tanaka:1994ay}. 

In a 2HDM of type II (with MSSM like Higgs potential) the only free additional parameters are $\tan\beta=v_u/v_d$ (the ratio of the two vacuum expectation values) and the charged Higgs mass $m_{H^\pm}$ (the heavy CP even Higgs mass $m_{H^0}$ and the CP odd Higgs mass $m_{A^0}$ can be expressed in terms of the charged Higgs mass and differ only by electroweak corrections). In this setup the charged Higgs contribution to $B\to \tau\nu$ interferes necessarily destructively with the SM contribution\cite{Hou:1992sy}. Thus, an enhancement of $\cal B\left[B\to \tau\nu\right]$ is only possible if the absolute value of the charged Higgs contribution is bigger than two times the SM one\footnote{Another possibility to explain $B\to \tau\nu$ is the introduction of a right-handed $W$-coupling \cite{Crivellin:2009sd}.}. Furthermore, a 2HDM of type II cannot explain $\cal{R}(D)$ and $\cal{R}(D^*)$ simultaneously \cite{BaBar:2012xj}.


\begin{figure}[t]
\centering
\includegraphics[width=0.3\textwidth]{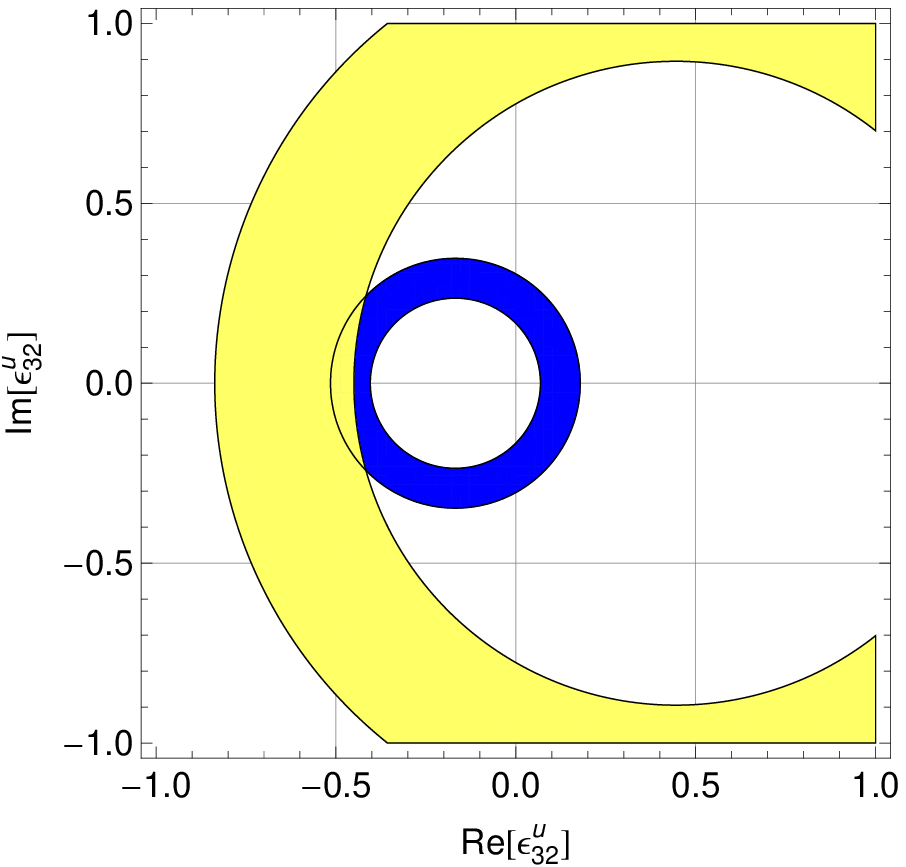}
\includegraphics[width=0.31\textwidth]{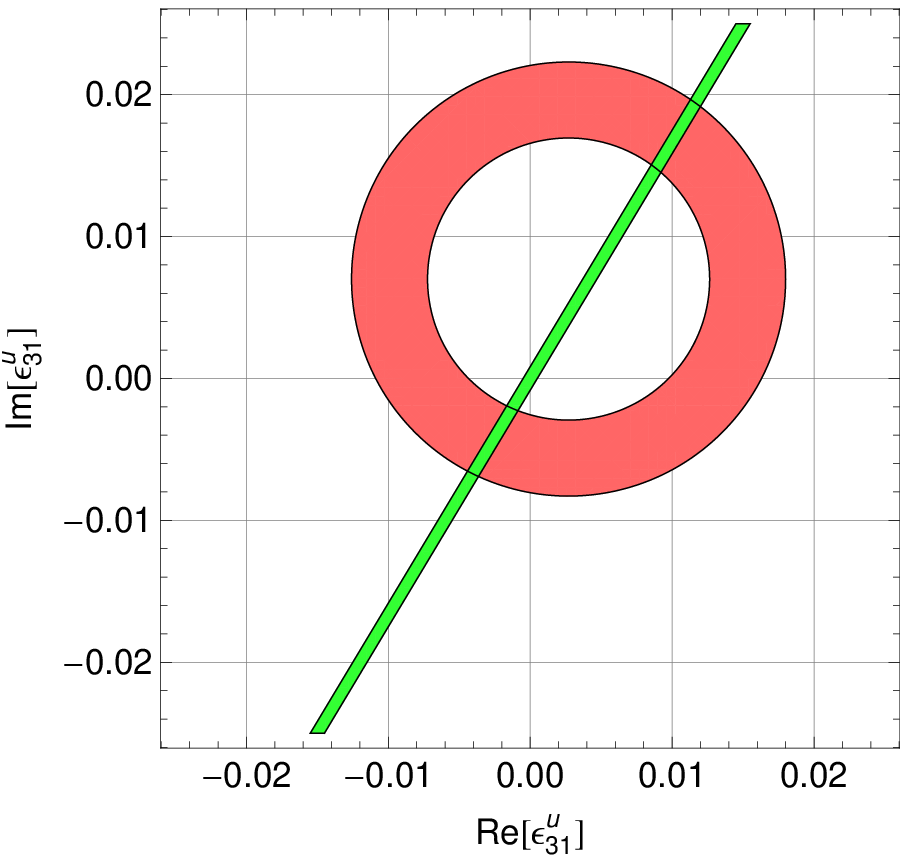}
\includegraphics[width=0.31\textwidth]{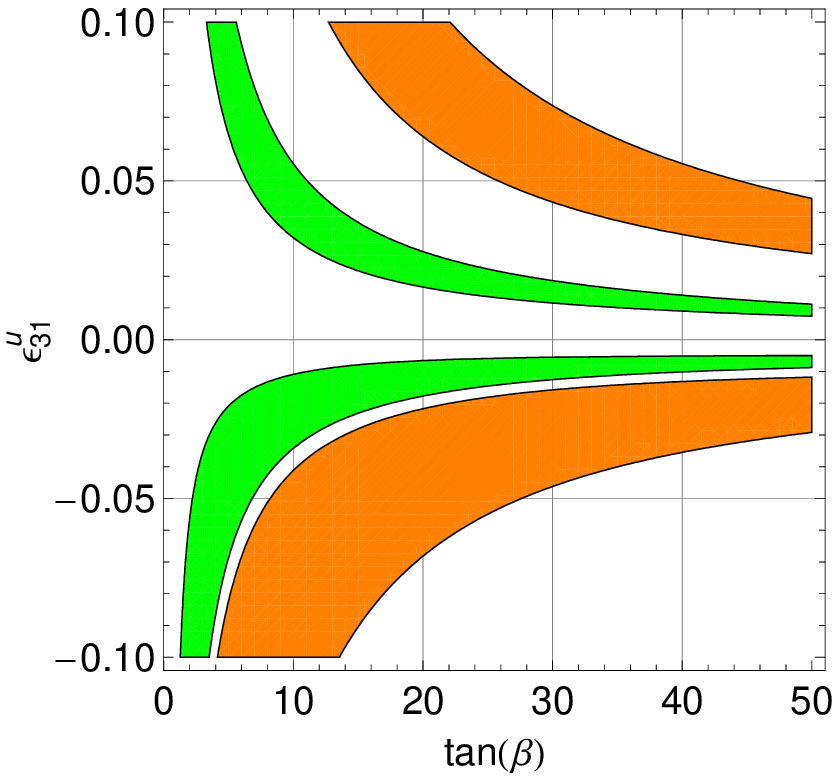}
\caption{Left: Allowed regions in the complex $\epsilon^u_{32}$-plane from $\cal{R}(D)$ (blue) and $\cal{R}(D^*)$ (yellow) for $\tan\beta=50$ and $m_H=500$~GeV. Middle:  Allowed regions in the complex $\epsilon^u_{31}$-plane from $B\to \tau\nu$. Right:  Allowed regions in the $\tan\beta$--$\epsilon^u_{31}$ plane from $B\to \tau\nu$ for real values of $\epsilon^u_{31}$ and $m_H=400$~GeV (green), $m_H=800$~GeV (orange). The scaling of the allowed region for $\epsilon^u_{32}$ with $\tan\beta$ and $m_H$ is the same as for $\epsilon^u_{31}$. $\epsilon^u_{32}$ and $\epsilon^u_{31}$ are given at the matching scale $m_H$. \label{2HDMIII}}
\end{figure}

As discussed in the last section we have much more free parameters ($\epsilon^q_{ij}$) in the 2HDM of type III which can in principle affect the tauonic $B$ decays. However, we found that all $\epsilon^d_{ij}$ are stringently constrained from FCNC processes in the down sector. Thus, they cannot have any significant impact on the decays we are interested in, and therefore we are left with~$\epsilon^d_{33}$. Concerning the elements $\epsilon^u_{ij}$ only $\epsilon^u_{31}$ ($\epsilon^u_{32}$) significantly effects $B\to \tau\nu$ ($\cal{R}(D)$ and $\cal{R}(D^*)$) without any suppression by small CKM elemets. Furthermore, since flavor-changing top-to-up (or charm) transitions are not measured with sufficient accuracy, we can only constrain these elements from charged Higgs-induced FCNCs in the down sector. However, since in this case an up (charm) quark always propagates inside the loop, the contribution is suppressed by the small Yukawa couplings of the up-down-Higgs (charm-strange-Higgs) vertex involved in the corresponding diagrams. Thus, the constraints from FCNC processes are weak, and $\epsilon^u_{32,31}$ can be sizable. Of course, the lower bounds on the charged Higgs mass for a 2HDM of type II from $b\to s\gamma$ of 380~GeV \cite{Hermann:2012fc} must still be respected by our model (unless $\epsilon^u_{23}$ generates a destructively interfering contribution), and also the results from direct searches at the LHC for $H^0,A^0\to\tau^+\tau^-$ \cite{Chatrchyan:2012vp} are in principle unchanged (if $\epsilon^\ell_{33}$ is not too large). 

Indeed, it turns out that by using $\epsilon^u_{32,31}$ we can explain $\cal{R}(D^*)$ and $\cal{R}(D)$ simultaneously which is not possible using $\epsilon^d_{33}$ alone. In Fig.~\ref{2HDMIII} we see the allowed region in the complex $\epsilon^u_{32}$-plane, which gives the correct values for $\cal{R}(D)$ and $\cal{R}(D^*)$ within the $1\, \sigma$ uncertainties for $\tan\beta=50$ and $M_H=500$~GeV. Similarly, $B\to \tau\nu$ can be explained by using $\epsilon^u_{31}$.

\section{Effective Higgs Vertices in the MSSM}
\label{sec:MSSM-2HDM}

In this section we discuss the matching of the MSSM on the 2HDM considering the Yukawa sector \footnote{For a discussion in MFV see for example \cite{Buras:2002vd,D'Ambrosio:2002ex}} but neglecting loop-corrections to the Higgs potential which to not lead to enhanced relations among parameters, i.e. the corrections can be reabsorbed by a redefinition of parameters \cite{Gorbahn:2009pp}. This means our goal is to express the parameters $\epsilon^q_{ij}$ in \eq{L-Y-FCNC} in terms of MSSM parameters. At tree-level, the MSSM is a 2HDM of type II but at the loop-level, the Peccei-Quinn symmetry of the Yukawa sector is broken by terms proportional to the higgsino mass parameter $\mu$ (or non-holomorphic $A^\prime$ terms) which then generates the non-holomorphic couplings $\epsilon^q_{ij}$. 

In the MSSM there is a one-to-one correspondence between Higgs-quark-quark couplings and chirality changing quark self-energies (in the decoupling limit\footnote{The non-decoupling corrections are found to be very small \cite{Crivellin:2010er}.}): The Higgs-quark-quark coupling can be obtained by dividing the expression for the self-energy by the vev of the corresponding Higgs field. 

Let us denote the contribution of the quark self-energy with squarks and gluinos to the operator $\overline{q}_f P_R q_i$ by $C_{f i }^{q\,LR}$. It is important to note that this Wilson coefficient is linear in $\Delta^{q\,LR}$, the off-diagonal element of the squark mass matrix connecting left-handed and right-handed squarks. For down squarks we have
\begin{equation}
	\Delta^{d\,LR}_{ij}=-v_d A^d_{ij}-v_u \mu Y^{d_i}\delta_{ij}\,,
\end{equation}
where the term $v_d A^d_{ij}$ originates from a coupling to $H^d$ while the term $v_u \mu Y^{d_i}$ stems from a coupling to $H^u$ (and similarly for up-squarks). Thus we denote the piece of $\hat C_{f i }^{d\,LR}$ involving the $A$-term by $\hat C^{d\,LR}_{fi\,A}$ and the piece containing $v_u \mu Y^{d_i}$ by $\hat C^{\prime\, d\,LR}_{fi}$. We now define 
\begin{equation}
\renewcommand{\arraystretch}{2}
\begin{array}{l}
   \hat E^d_{fi}\,=\,\dfrac{\hat C^{d\,LR}_{fi\,A}}{v_d}\,,\hspace{0.5cm} 
   \hat E^{\prime d}_{fi}\,=\,\dfrac{\hat C^{\prime\, d\,LR}_{fi}}{v_u}\,,\hspace{0.5cm} 
   \hat E^u_{fi}\,=\,\dfrac{\hat C^{u\,LR}_{fi\,A}}{v_u}\,,\hspace{0.5cm} 
   \hat E^{\prime u}_{fi}\,=\,\dfrac{\hat C^{\prime\, u\,LR}_{fi}}{v_d}\,,
   \end{array}
   \label{E-Sigma}
\end{equation}
where the parameters $\hat E_{fi}^{q}$ ($\hat E_{fi}^{\prime q}$) correspond to (non-)holomorphic Higgs-quark couplings. With these conventions, the couplings  $\epsilon^q_{ij}$ of the 2HDM in \eq{L-Y-FCNC} can be related to MSSM parameters
\begin{eqnarray}
\renewcommand{\arraystretch}{2.0}
\begin{array}{l}
 \epsilon_{fi}^{q}  = \hat E_{fi}^{\prime q}  - \left( 
 {\begin{array}{*{20}c}
   0 & 
   {\hat E_{22}^{\prime q} \frac{\hat C_{12}^{q\,LR}}{m_{q_2}} } 
   & \hat E_{33}^{\prime q} \left(  \frac{\hat C_{13}^{q\,LR}}{m_{q_3}}  - \frac{\hat C _{12}^{q\;LR}}{m_{q_2}} \frac{\hat C_{23}^{q\,LR}}{m_{q_3}}  \right)  \\
   {\hat E_{22}^{\prime q} \frac{\hat C _{21}^{q\;LR}}{m_{q_2}} } 
   & 0 
   & {\hat E_{33}^{\prime q} \frac{\hat C _{23}^{q\;LR}}{m_{q_3}}}  \\
   {\hat E_{33}^{\prime q} \left( {\frac{\hat C _{31}^{q\,LR}}{m_{q_3}}  - \frac{\hat C _{32}^{q\;LR}}{m_{q_3}} \frac{\hat C _{21}^{q\,LR}}{m_{q_2}} } \right)} 
   & {\hat E_{33}^{\prime q} \frac{\hat C _{32}^{q\,LR}}{m_{q_3}} } 
   & 0  \\
\end{array}} \right)_{fi}  \,. 
 \end{array}\nonumber
\label{Etilde}
\end{eqnarray}

\begin{figure}
\centering
\includegraphics[width=0.49\textwidth]{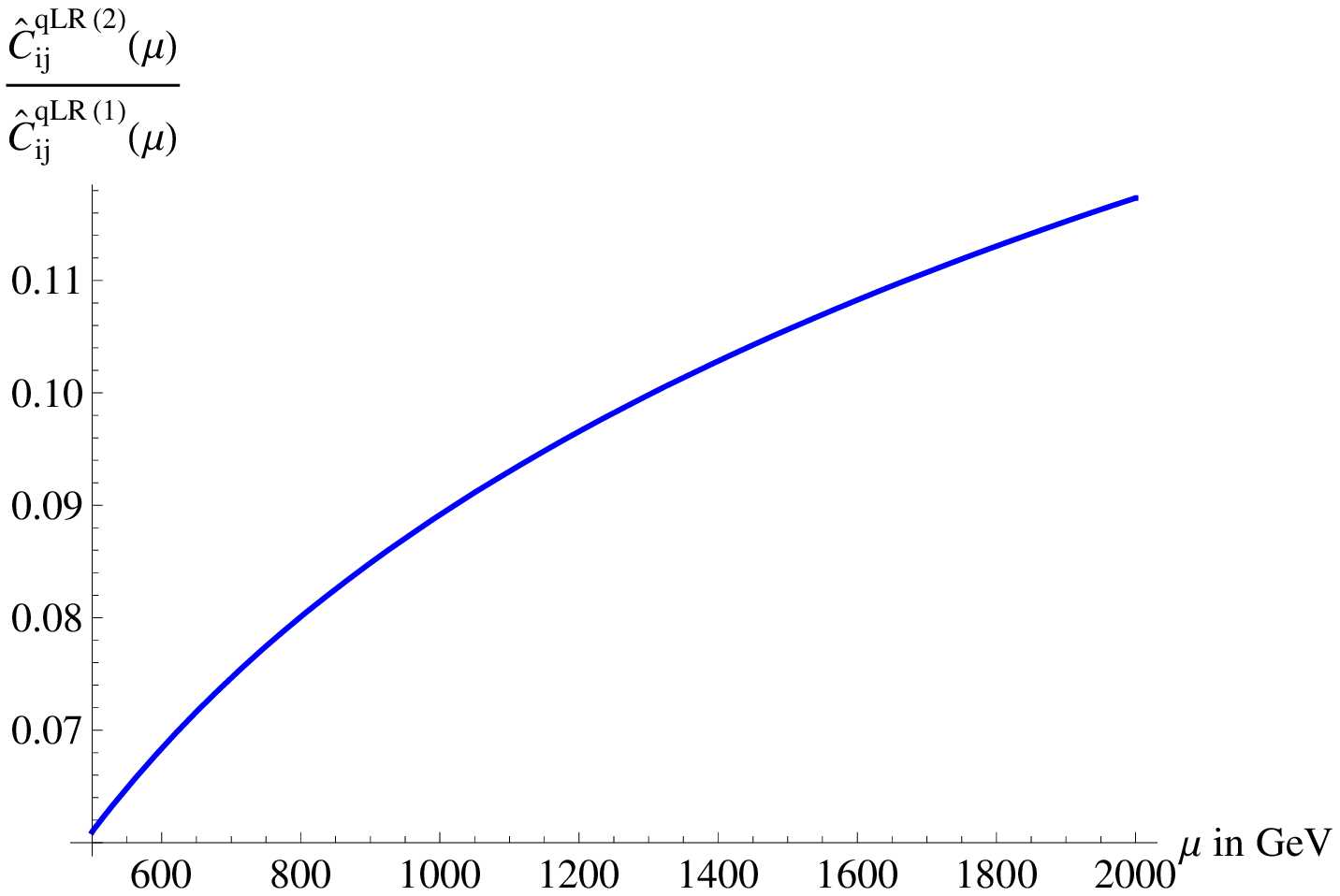}
\includegraphics[width=0.49\textwidth]{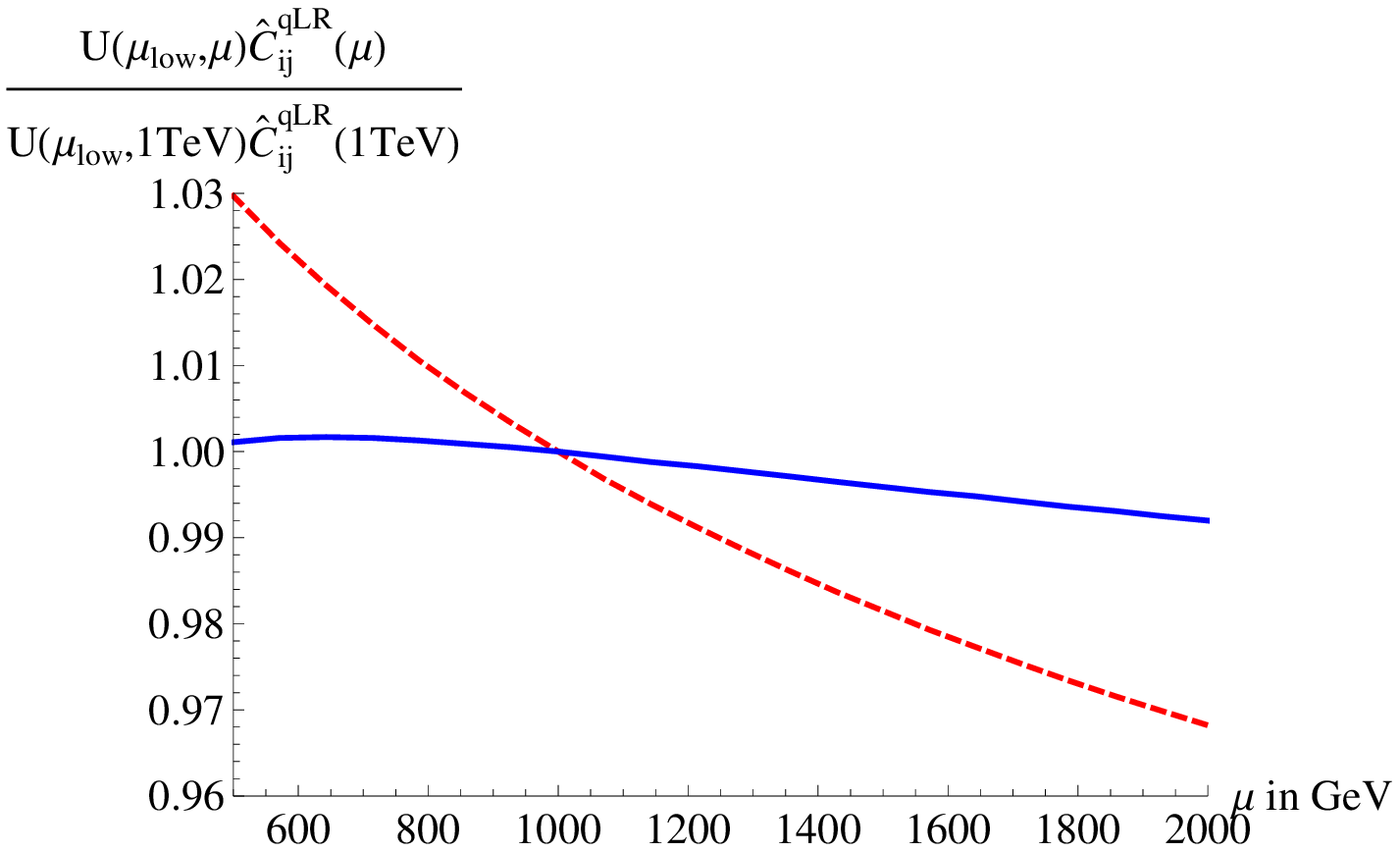}
\caption{Left:Relative importance of the two-loop corrections as a function of the matching scale $\mu$. We see that the two-loop contribution is approximately +9\% of the one-loop contribution for $\mu=M_{\rm SUSY}=1\, {\rm TeV}$.\newline
Right: Dependence on the matching scale $\mu$ of the one-loop and
  two-loop result for $\hat C _{fi}^{q\,LR}(\mu_{\rm low})$, using $M_{\rm SUSY}=1$~TeV and $\mu_{\rm low}=m_W$. Red (dashed): matching done at LO;  blue (darkest):  matching done at NLO matching.  As expected, the
  matching scale dependence is significantly reduced. For the one-loop result, $\hat  C _{fi}^{q\,LR}$ is understood to be $C_{fi}^{q\,LR\,(1)}$ (see text).
  \label{mu-abhaengigkeit}}
\end{figure}

In the matching of the MSSM on the 2HDM one can as a by product also determine the Yukawa couplings of the MSSM superpotential which is important for the study of Yukawa coupling unification in supersymmetric GUTs. Due to this importance of the chirality changing self-energies we calculated them (and thus also $\hat C^{q\,LR}_{ij}$) at the two loop-level in Ref.~\cite{Crivellin:2012zz}\footnote{For eariler results in the MSSM with MFV see \cite{Noth:2010jy}}. The result is a reduction of the matching scale dependence (see right plot of Fig.~\ref{mu-abhaengigkeit}) while at the same time, the one-loop contributions are enhanced by a relative effect of 9\% (see left plot of Fig.~\ref{mu-abhaengigkeit}). For a numerical analysis also the LO chargino and neutralino contributions should be included by using the results of Ref.~\cite{Crivellin:2010er}.

Concerning the tauonic $B$-decays discussed in the last section, the size of the quantities $\epsilon^u_{32,31}$ that can be generated via loops in the MSSM is too small to give a sizable effect.

\acknowledgments

I thank the organizers, especially Gino Isidori, for the invitation to "Les Rencontres de Physique de la Vallée d'Aoste". This work is supported by a Marie Curie Intra-European Fellowship of the European Community's 7th Framework Programme under contract number (PIEF-GA-2012-326948).


\begin{thebibliography}{99}

\bibitem{Crivellin:2012ye}
  A.~Crivellin, C.~Greub and A.~Kokulu,
  ``Explaining $B\to D\tau\nu$, $B\to D^*\tau\nu$ and $B\to \tau\nu$ in a 2HDM of type III,''  Phys.\ Rev.\ D {\bf 86} (2012) 054014  [arXiv:1206.2634 [hep-ph]].  
	
\bibitem{Crivellin:2013wna}
  A.~Crivellin, A.~Kokulu and C.~Greub,
  ``Flavor-phenomenology of two-Higgs-doublet models with generic Yukawa structure,''  Phys.\ Rev.\ D {\bf 87} (2013) 9,  094031  [arXiv:1303.5877 [hep-ph]].  


\bibitem{Crivellin:2012zz}
  A.~Crivellin and C.~Greub,
  ``Two-loop SQCD corrections to Higgs-quark-quark couplings in the generic MSSM,''  arXiv:1210.7453 [hep-ph].  
  
\bibitem{Lee:1973iz}
  T.~D.~Lee,
  ``A Theory of Spontaneous T Violation,''  Phys.\ Rev.\ D {\bf 8} (1973) 1226.  

\bibitem{Hamzaoui:1998nu}
  C.~Hamzaoui, M.~Pospelov and M.~Toharia,
  ``Higgs mediated FCNC in supersymmetric models with large $\tan \beta$,''  Phys.\ Rev.\ D {\bf 59} (1999) 095005  [hep-ph/9807350].\\  
  K.~S.~Babu and C.~F.~Kolda,
  ``Higgs mediated $B^0 \to \mu^{+} \mu^{-}$ in minimal supersymmetry,''  Phys.\ Rev.\ Lett.\  {\bf 84} (2000) 228  [hep-ph/9909476].\\  
  G.~Isidori and A.~Retico,
  ``$B_{s,d} \to \ell^{+} \ell^{-}$ and $K_{L} \to \ell^{+} \ell^{-}$ in SUSY models with nonminimal sources of flavor mixing,''  JHEP {\bf 0209} (2002) 063  [hep-ph/0208159].  

\bibitem{Crivellin:2010er}
  A.~Crivellin,
  ``Effective Higgs Vertices in the generic MSSM,''  Phys.\ Rev.\ D {\bf 83} (2011) 056001  [arXiv:1012.4840 [hep-ph]].  
  A.~Crivellin, L.~Hofer and J.~Rosiek,
  ``Complete resummation of chirally-enhanced loop-effects in the MSSM with non-minimal sources of flavor-violation,''  JHEP {\bf 1107} (2011) 017  [arXiv:1103.4272 [hep-ph]].  

\bibitem{Sher:1991km}
  M.~Sher and Y.~Yuan,
  ``Rare B decays, rare tau decays and grand unification,''  Phys.\ Rev.\ D {\bf 44} (1991) 1461.  

\bibitem{CMS}
  S.~Chatrchyan {\it et al.}  [CMS Collaboration],
  ``Search for supersymmetry in hadronic final states using MT2 in $pp$ collisions at $\sqrt{s} = 7$ TeV,''
  JHEP {\bf 1210} (2012) 018
  [arXiv:1207.1798 [hep-ex]].

\bibitem{Crivellin:2011ba}
  A.~Crivellin and L.~Mercolli,
  ``$B -> X_d \gamma$ and constraints on new physics,''  Phys.\ Rev.\ D {\bf 84}  (2011) 114005   [arXiv:1106.5499 [hep-ph]].  


\bibitem{BaBar:2012xj}
  J.~P.~Lees {\it et al.}  [BaBar Collaboration],
  ``Evidence for an excess of $\bar{B} \to D^{(*)} \tau^-\bar{\nu}_\tau$ decays,''  Phys.\ Rev.\ Lett.\  {\bf 109} (2012) 101802 [arXiv:1205.5442 [hep-ex]].  

\bibitem{Charles:2004jd}
  J.~Charles {\it et al.}  [CKMfitter Group Collaboration],
  ``CP violation and the CKM matrix: Assessing the impact of the asymmetric $B$ factories,''  Eur.\ Phys.\ J.\ C {\bf 41} (2005) 1  [hep-ph/0406184].  

\bibitem{Hou:1992sy}
  W.~-S.~Hou,
  ``Enhanced charged Higgs boson effects in $B \to \tau\nu$ and $B \to \tau\nu +X$,''  Phys.\ Rev.\ D {\bf 48} (1993) 2342.  

\bibitem{Tanaka:1994ay}
  M.~Tanaka,
  ``Charged Higgs effects on exclusive semitauonic $B$ decays,''  Z.\ Phys.\ C {\bf 67} (1995) 321  [hep-ph/9411405].  

\bibitem{Crivellin:2009sd}
  A.~Crivellin,
  ``Effects of right-handed charged currents on the determinations of $|V_{ub}|$ and $|V_{cb}|$,''  Phys.\ Rev.\ D {\bf 81} (2010) 031301  [arXiv:0907.2461 [hep-ph]].  

\bibitem{Hermann:2012fc}
  T.~Hermann, M.~Misiak and M.~Steinhauser,
  ``$\bar{B}\to X_s \gamma$ in the Two Higgs Doublet Model up to Next-to-Next-to-Leading Order in QCD,''  JHEP {\bf 1211} (2012) 036  [arXiv:1208.2788 [hep-ph]].  
  
\bibitem{Chatrchyan:2012vp}
  S.~Chatrchyan {\it et al.}  [CMS Collaboration],
  ``Search for neutral Higgs bosons decaying to tau pairs in pp collisions at sqrt(s)=7 TeV,''  Phys.\ Lett.\ B {\bf 713} (2012) 68  [arXiv:1202.4083 [hep-ex]].  
	
\bibitem{Buras:2002vd}
  A.~J.~Buras, P.~H.~Chankowski, J.~Rosiek and L.~Slawianowska,
  ``$\Delta M_{d,s}, B^0{d,s} \to \mu^{+} \mu^{-}$ and $B \to X_{s} \gamma$ in supersymmetry at large $\tan\beta$,''  Nucl.\ Phys.\ B {\bf 659} (2003) 3  [hep-ph/0210145].  
\bibitem{D'Ambrosio:2002ex}
  G.~D'Ambrosio, G.~F.~Giudice, G.~Isidori and A.~Strumia,
  ``Minimal flavor violation: An Effective field theory approach,''  Nucl.\ Phys.\ B {\bf 645} (2002) 155  [hep-ph/0207036].  

\bibitem{Gorbahn:2009pp}
  Y.~Okada, M.~Yamaguchi and T.~Yanagida,
  ``Upper bound of the lightest Higgs boson mass in the minimal supersymmetric standard model,''  Prog.\ Theor.\ Phys.\  {\bf 85} (1991) 1.  
  P.~H.~Chankowski, S.~Pokorski and J.~Rosiek,
  ``Complete on-shell renormalization scheme for the minimal supersymmetric Higgs sector,''  Nucl.\ Phys.\ B {\bf 423} (1994) 437  [hep-ph/9303309].  
  A.~Dabelstein,
  ``The One loop renormalization of the MSSM Higgs sector and its application to the neutral scalar Higgs masses,''  Z.\ Phys.\ C {\bf 67} (1995) 495  [hep-ph/9409375].  
  M.~Gorbahn, S.~Jager, U.~Nierste and S.~Trine,
  ``The supersymmetric Higgs sector and $B-\bar{B}$ mixing for large tan $\beta$,''  Phys.\ Rev.\ D {\bf 84}  (2011) 034030   [arXiv:0901.2065 [hep-ph]].  

\bibitem{Noth:2010jy}
  D.~Noth and M.~Spira,
  ``Supersymmetric Higgs Yukawa Couplings to Bottom Quarks at next-to-next-to-leading Order,''  JHEP {\bf 1106} (2011) 084  [arXiv:1001.1935 [hep-ph]].  
  A.~Bauer, L.~Mihaila and J.~Salomon,
  ``Matching coefficients for alpha(s) and m(b) to $O(\alpha_s^2)$ in the MSSM,''  JHEP {\bf 0902} (2009) 037  [arXiv:0810.5101 [hep-ph]].  
  A.~Bednyakov, A.~Onishchenko, V.~Velizhanin and O.~Veretin,
  ``Two loop $O(\alpha_s^2)$ MSSM corrections to the pole masses of heavy quarks,''  Eur.\ Phys.\ J.\ C {\bf 29} (2003) 87  [hep-ph/0210258].  

	


\end{thebibliography}
\end{document}